\documentclass[twocolumn,showpacs,amsmath,amssymb,nofootinbib]{revtex4}

\usepackage{bm}
\newtheorem{theorem}{Theorem}
\newtheorem{lemma}[theorem]{Lemma}

\begin{document}

\title{Lifting Bell inequalities}
\author{Stefano Pironio}
\email{spironio@caltech.edu}
\affiliation{Institute for Quantum Information,\\ California Institute of Technology,\\ Pasadena, CA 91125, USA}
\date{June 17, 2005}
\begin{abstract}
A Bell inequality defined for a specific experimental configuration can always be extended to a situation involving more observers, measurement settings or measurement outcomes. In this article, such ``liftings" of Bell inequalities are studied. It is shown that if the original inequality defines a facet of the polytope of local joint outcome probabilities then the lifted one also defines a facet of the more complex polytope.
\end{abstract}
\maketitle

\section{Introduction}
In a typical Bell experiment, two or more entangled particles are distributed to separate observers. Each observer measures on his particle one from a set of possible observables and obtains some outcome.  One of the most striking features of quantum mechanics is that the resulting joint outcome probabilities can violate a Bell inequality \cite{bel64}, indicating that quantum mechanics is not, in Bell's terminology, locally causal. This prediction has been confirmed, up to some loopholes, in numerous laboratory experiments \cite{asp99,tw01}. The implications of nonlocality for our fundamental description of nature \cite{bel87,cm89} have long been discussed; more recently, nonlocality has also acquired a significance in quantum information science \cite{eke91,ags03,bhk04,asw02,bra03,bzp04,blm05}. From this perspective, being able to decide whether a joint probability distribution can be reproduced with classical randomness only, or whether entanglement is necessary, is an important issue.

For a given number of observers, measurement settings, and measurement outcomes, the set of joint probabilities accessible to locally causal theories is a convex polytope \cite{ww01b}. It is therefore completely characterized by a finite number of linear inequalities that these probabilities must satisfy --- that is, by a finite number of Bell inequalities. Each of these inequalities corresponds to a \emph{facet} of the local polytope. Note, however, that not every Bell inequality represents a facet. Facet inequalities are the ones which characterize precisely the border between the local and the nonlocal region. They form a minimal and complete set of Bell inequalities.

In the simple situation where they are only two observers, two measurement choices, and two outcomes per measurement, all the facet inequalities are known \cite{fro81,fin82}: up to permutation of the outcomes, they correspond to the Clauser-Horne-Shimony-Holt (CHSH) inequality \cite{chs69}. Beyond this, little is known. It is in principle possible to obtain all the facet inequalities of an arbitrary Bell polytope using specific algorithms. In practice this only allows one to extend the range of solved cases to a few more observers, measurements or outcomes \cite{ps01,cg04}, as these algorithms are excessively time-consuming. 
The problem of listing all facet inequalities has in fact been demonstrated to be NP-complete \cite{aii04}; it is therefore unlikely that it could be solved in full generality. Discouraging as this result may seem, it nevertheless leaves open several possibilities. First, complete sets of facet inequalities may be obtained for particular classes of Bell polytopes or for simplified versions of them. For instance, in the case where ``full correlation functions" are considered instead of complete joint probability distributions, all facet inequalities are known for Bell scenarios consisting of an arbitrary number of parties with two measurement choices and two outcomes \cite{ww01,zb02}. Second, in more complicated situations it may still be possible to obtain partial lists of facets. For instance, families of facet inequalities are known for arbitrary number of measurements \cite{aii04} or outcomes \cite{mas03}. 

Further progress in the derivation of Bell inequalities would certainly benefit from a better characterization of the general properties of Bell polytopes. This is the motivation behind the present article. The question that we will investigate is how, and to what extent, the facial structure of a Bell polytope determines the facial structure of more complex polytopes. More specifically consider a bipartite Bell experiment characterized by the probability $p_{k_1k_2|j_1j_2}$ for the first observer to obtain outcome $k_1$ and for the second one to obtain outcome $k_2$, given that the first observer measures $j_1$ and the second one $j_2$. Suppose that each observer chooses one from two dichotomic observables, that is, $k_1,k_2\in \{1,2\}$ and $j_1,j_2\in \{1,2\}$. A necessary condition for this experiment to be reproducible by a local model is that the joint probabilities satisfy the CHSH inequality
\begin{eqnarray}\label{chsh}
&p_{11|11}+p_{11|12}+p_{11|21}-p_{11|22}&\nonumber \\
+&p_{22|11}+p_{22|12}+p_{22|21}-p_{22|22}&\geq 0\,.
\end{eqnarray}
Although this inequality is defined for the specific Bell scenario that we have just described, it also constrains the set of local joint probabilities involving more observers, measurements, and outcomes. Indeed, as was noted by Peres \cite{per99} there are obvious ways to extend Bell inequalities to more complex situations, or to \emph{lift} them following the terminology of polytope theory. As an illustration, let us consider the following three possible extensions of our CHSH scenario.

\emph{(i) More observers.} Consider a tripartite Bell experiment with joint probability distribution $p_{k_1k_2k_3|j_1j_2j_3}$, where $k_1,k_2,k_3\in \{1,2\}$ and $j_1,j_2,j_3\in \{1,2\}$. A necessary condition for this tripartite distribution to be local is that the probabilities $\widetilde p_{k_1k_2|j_1j_2}$ for the first two observers to measure $j_1$ and $j_2$ and to obtain outcomes $k_1$ and $k_2$ \emph{conditional} on the third observer measuring $j_3=1$ and obtaining $k_3=1$ satisfy the CHSH inequality. These conditional probabilities are given by $\widetilde p_{k_1k_2|j_1j_2}=p_{k_1k_21|j_1j_21}/p_{1_3|1_3}$, where the marginal $p_{1_3|1_3}=\sum_{k_1,k_2}p_{k_1k_21|j_1j_21}$ is independent of $j_1$ and $j_2$ by nosignaling\footnote{See Section \ref{dim}.}. Inserting these probabilities in (\ref{chsh}) and multiplying both side by $p_{1_3|1_3}$ leads to
\begin{eqnarray}\label{chshmo}
&p_{111|111}+p_{111|121}+p_{111|211}-p_{111|221}&\nonumber \\
+&p_{221|111}+p_{221|121}+p_{221|211}-p_{221|221}&\geq 0\,,
\end{eqnarray}
a natural extension of the CHSH inequality to three parties.

\emph{(ii) More measurements.} Consider our original bipartite Bell scenario, but assume that the second observer may choose between three different measurement settings $j_2\in\{1,2,3\}$. Clearly, a necessary condition for the corresponding joint distribution to be reproducible by a local model is that, when restricted to the probabilities involving $j_2\in\{1,2\}$, it satisfies the CHSH inequality. Therefore, inequality (\ref{chsh}) is, as such, a valid Bell inequality for this three-measurement scenario.

\emph{(iii) More outcomes.} Suppose now that the measurement apparatus of the second observer may output one out of three distinct values $k_2\in\{1,2,3\}$. Merging the outcomes $k_2=2$ and $k_2=3$, we obtain an effective two-outcomes distribution with probabilities $\widetilde p_{k_11|j_1j_2}=p_{k_11|j_1j_2}$ and $\widetilde p_{k_12|j_1j_2}=p_{k_12|j_1j_2}+p_{k_13|j_1j_2}$. The existence of a local model for the original distribution obviously implies a model for the coarse-grained one. Expressing the fact that the $\widetilde p_{k_1k_2|j_1j_2}$ should satisfy (\ref{chsh}), we thus deduce the following lifting
\begin{eqnarray}
&p_{11|11}+p_{11|12}+p_{11|21}-p_{11|22}&\nonumber \\
+&p_{22|11}+p_{22|12}+p_{22|21}-p_{22|22}&\nonumber \\
+&p_{23|11}+p_{23|12}+p_{23|21}-p_{23|22}&\geq 0
\end{eqnarray}
of the CHSH inequality to three outcomes.

These three examples can be combined and used sequentially to lift the CHSH inequality to an arbitrary number of observers, measurements, and outcomes. It is also straightforward to generalize them to other Bell inequalities than the CHSH one. How strong are the constraints on the joint probabilities obtained in this way? We will show that if the original inequality describes a facet of the original polytope, then the lifted one is also a facet of the more complex polytope. This implies, for instance, that the CHSH inequality is a facet of every Bell polytope since it is a facet of the simplest one.

This article is organized as follows. Section II introduces the concepts and notations that will be used in the remainder of the paper. In particular, we briefly review the definition of Bell polytopes and elementary notions of polytope theory. In Section III, we derive some basic properties of Bell polytopes that are necessary to prove our main results concerning the lifting of facet inequalities. These results are presented in Section IV. We conclude with a discussion and some open questions in Section V. 
 
\section{Definitions}
\subsection{Bell scenario}
Consider $n$ systems and assume that on each system $i$ a measurement $j\in\{1,\ldots,m_i\}$ is made, yielding an outcome $k\in\{1,\ldots, v_{ij}\}$. Note that the number of possible measurements $m_i$ may be different for each system $i$, and that the number of possible outcomes $v_{ij}$ may be different for each measurement $j$ on system $i$. Such a Bell scenario is thus characterized by the triple $(n,m,v)$ where $m=(m_1,\ldots,m_n)$ specifies the number of possible measurements per system, and where the table $v=\big[(v_{11},\ldots,v_{1m_1});\ldots;(v_{n1},\ldots,v_{nm_n})\big]$ specifies the number of possible outcomes per measurement on each system. When notations such as $(n,2,v)$ are used, it should be understood that $m_i=2$ for all $i$.

The joint probability of obtaining the outcomes $(k_1,\ldots,k_n)$ given the measurement settings $(j_1,\ldots,j_n)$ will be denoted $p_{k_1\ldots k_n|j_1 \ldots j_n}$. We will view these $t=\prod_{i=1}^{n}\left(\sum_{j=1}^{m_i}v_{ij}\right)$ probabilities as forming the components of a vector $p$ in $\mathbb{R}^t$.
For a given observer $i\in\{1,\ldots,n\}$, measurement $j\in\{1,\ldots,m_i\}$ and outcome $k\in\{1,\ldots,v_{ij}\}$, we will often be interested in the subset of the components of $p$ that have the indices $k_i$ and $j_i$ corresponding to observer $i$ fixed, and equal, respectively, to $k$ and $j$. In other words, we will be interested in the variables $p_{k_1\ldots k_{i-1}k\,k_{i+1}\ldots k_n|j_1\ldots j_{i-1}j\,j_{i+1}\ldots j_n}$. The restriction of $p$ to these components will be denoted $p(i,j,k)$.

\subsection{Bell polytopes}
The set $\mathcal{B}\subseteq \mathbb{R}^t$ of correlations reproducible within a locally causal model is the set of correlations $p$ satisfying
\begin{equation*}
p_{k_1\ldots k_n|j_1\ldots j_n}=\int\!\mathrm{d} \mu\, q(\mu) P(k_1|j_1,\mu)\ldots P(k_n|j_n,\mu)\,,
\end{equation*}
where $q(\mu)\geq 0$, $\int\!\mathrm{d}\mu\, q(\mu)=1$, and $P(k_i|j_i,\mu)$ is the probability of obtaining the measurement outcome $k_i$ given the setting $j_i$ and the hidden-variable $\mu$ \cite{bel64,bel87}.
From this definition it is easily deduced (see \cite{ww01b} for instance) that $p$ is generated by specifying probabilities for every assignment of one of the possible outcomes to each of the measurement settings. More precisely, let the table $\lambda=\big[(\lambda_{11},\ldots,\lambda_{1m_1});\ldots;(\lambda_{n1},\ldots,\lambda_{nm_n})\big]$ assign to each measurement $j$ on system $i$ the outcome $\lambda_{ij}$. The (finite) set of all such possible assigmenents will be denoted $\Lambda$. Let 
\begin{equation}\label{defdetvect}
p^\lambda_{k_1\ldots k_n|j_1\ldots j_n}=\left\{\begin{array}{ll}1 &\text{if }\lambda_{1j_1}=k_1,\ldots,\lambda_{nj_n}=k_n\\ 0&\mbox{otherwise}\end{array}\right.
\end{equation}
be the deterministic vector corresponding to the assignment $\lambda$. Then 
\begin{equation}\label{localpolya}
\mathcal{B}=\{p\in\mathbb{R}^t\mid p=\sum_{\lambda\in\Lambda} q_\lambda\, p^\lambda,\, q_\lambda\geq0,\, \sum_{\lambda\in\Lambda} q_\lambda=1\}\,.
\end{equation}
The set $\mathcal{B}$ of local correlations is thus the convex hull of a finite number of points, i.e., it is a polytope. The deterministic vectors $\{p^\lambda|\lambda\in\Lambda\}$ form the extreme points of this polytope.

\subsection{Notions of polytope theory}\label{polrev}
We review in this section some elementary notions of polytope theory. For more detailed introductions, see \cite{nw88,sch89,zie95}.

The points $p_1,\ldots,p_n$ in $\mathbb{R}^t$ are said to be affinely independent if the unique solution to $\sum_i \mu_ip_i=0$, $\sum_i\mu_i=0$ is $\mu_i=0$ for all $i$, or equivalently, if the points $p_2-p_1,\ldots,p_n-p_1$ are linearly independent. They are affinely dependent otherwise. The affine hull of a set of points is the set of all their affine combinations. An affine set has dimension $D$, if the maximum number of affinely independent points it contains is $D+1$.

Let $\mathcal{B}\subseteq \mathbb{R}^t$ be a polytope defined as in (\ref{localpolya}). Let $(b,b_0)\in\mathbb{R}^{t+1}$ define the inequality $b\cdot p\geq b_0$. If this inequality is satisfied for all $p\in\mathcal{B}$, it is called a valid inequality for the polytope $\mathcal{B}$, or a Bell inequality in the context of Bell polytopes. Note that to check whether an inequality is a valid inequality, it is sufficient, by convexity, to check whether it is satisfied by the extreme points $\{p^\lambda|\lambda\in\Lambda\}$. Given the valid inequality $b\cdot p\geq b_0$, the set $F=\{p\in \mathcal{B}\mid b\cdot p=b_0\}$ is called a face of $\mathcal{B}$ and the inequality is said to support $F$. If $F\neq\emptyset$ and $F\neq\mathcal{B}$, it is a proper face. The dimension of $F$ is the dimension of its affine hull. Proper faces clearly satisfy $\dim F\leq \dim \mathcal{B}-1$. Proper faces of maximal dimension are called facets. An inequality $b\cdot p\geq b_0$ thus supports a facet of $\mathcal{B}$ if and only if $\dim \mathcal{B}$ affinely independent of $\mathcal{B}$ satisfy it with equality.

A fundamental result in polyhedral theory, known as Minkowski-Weyl's theorem, states that a polytope represented as the convex hull of a finite number of points, as in (\ref{localpolya}), can equivalently be represented as the intersection of finitely many half-spaces:
\begin{equation}\label{localpolyb}
\mathcal{B}=\{p\in\mathbb{R}^t\mid b^i\cdot p\geq b^i_0,\, \mbox{for all } i\in I\}\,,
\end{equation}
where $\{b^i\cdot p\geq b^i_0,\,i\in I\}$ is a finite set of inequalities. The inequalities supporting facets of $\mathcal{B}$ provide a minimal set of such inequalities\footnote{Note that if $\mathcal{B}\subseteq \mathbb{R}^t$ is not full dimensional, that is if $\dim\mathcal{B}<t$, then \emph{equality constraints} describing the affine hull of $\mathcal{B}$ must also be included in the above description.}. In particular, any valid inequality for $\mathcal{B}$ can be derived from the facet inequalities. 

Given a Bell scenario $(n,m,v)$, the task of finding all the Bell inequalities is thus the problem of finding all the facets of the convex polytope $\mathcal{B}(n,m,v)$ defined by (\ref{defdetvect}) and (\ref{localpolya}). This connection between the search for optimal Bell inequalities and polyhedral geometry was observed by different authors \cite{fro81,gm84,pit89,per99}. For discussions on the complexity of this facet enumeration task see \cite{pit91,aii04}. For the instances for which this problem has been partially or completely solved, see \cite{fro81,fin82,ps01,ww01,zb02,mas03,sli03,cg04,aii04,lpz04}. 

\section{Basic properties of Bell polytopes}\label{bp}
\subsection{Affine hull}\label{dim}
Local correlations $p\in\mathcal{B}$ satisfy the following equality constraints:\\
\emph{The normalization conditions}
\begin{equation}\label{multinorma}
\sum_{k_1\ldots k_n} p_{k_1 \ldots k_n|j_1 \ldots j_n}=1
\end{equation}
for all $j_1,\ldots,j_n$;\\
\emph{and the nosignaling conditions}
\begin{equation}\label{multinosig}
\sum_{k_{i}}p_{k_1\ldots k_{i}\ldots k_n|j_1\ldots j_{i}\ldots j_n}=\sum_{k_{i}}p_{k_1\ldots k_{i}\ldots k_n|j_1\ldots j'_{i}\ldots j_n}
\end{equation}
for all $i$, $k_1,\ldots k_{i-1},k_{i+1},\ldots,k_n$ and $j_1,\ldots j_{i-1},j_{i},j'_{i},\linebreak[4]j_{i+1},\ldots,j_n$.

The nosignaling conditions imply that for each subset $\{i_1,\ldots,i_q\}$ of size $q$ of the observers, the $q$-marginals $p_{k_{i_1}\ldots k_{i_q}|j_{i_1}\ldots j_{i_q}}=\sum_{k_{i_{q+1}}}\ldots\sum_{k_{i_n}}p_{k_1\ldots k_n|j_1\ldots j_n}$ are well-defined, that is, are independent of the precise value of the measurement settings $j_{i_{q+1}}\ldots j_{i_n}$.

The two conditions (\ref{multinorma}) and (\ref{multinosig}) also imply that the polytope $\mathcal{B}$ is not full dimensional in $\mathbb{R}^t$, i.e., it is contained in an affine subspace. The following theorem generalizes results given in \cite{mas03} and \cite{aii04}.
\begin{theorem}\label{localdim}
The constraints (\ref{multinorma}) and (\ref{multinosig}) fully determine the affine hull of $\mathcal{B}$ and
\begin{equation}
\dim\mathcal{B}=\prod_{i=1}^n\left(\sum_{j=1}^{m_i}\left(v_{ij}-1\right)+1\right)-1\,.
\end{equation}
\end{theorem}
\noindent\emph{Proof.}
Consider the marginals $p_{k_{i_1}\ldots k_{i_q}|j_{i_1}\ldots j_{i_q}}$ as defined above for all possible subsets $\{i_1,\ldots,i_q\}$ of size $q$, and for all $q=1,\ldots,n$. Of these marginals retain only the ones such that $k_{i}\neq 1$ for all $i$ $\in\{i_1,\ldots,i_q\}$. These probabilities define in total $D=\prod_{i=1}^n\Big(\sum_{j=1}^{m_i}(v_{ij}-1)+1\Big)-1$ numbers. It is straightforward to check that their knowledge is sufficient to reconstruct, using the normalization and nosignaling conditions, the original $p_{k_1 \ldots k_n|j_1 \ldots j_n}$. This implies that the affine subspace defined by (\ref{multinorma}) and (\ref{multinosig}) is of dimension $\leq D$.

Let us now show that $\dim\mathcal{B}\geq D$, or equivalently that $\mathcal{B}$ contains $D+1$ affinely independent points. For this, note that the definition (\ref{defdetvect}) implies that an extreme point $p^\lambda$ can be written as the product $p^\lambda_{k_1\ldots k_n|j_1\ldots j_n}=p^\lambda_{k_1|j_1}\ldots p^\lambda_{k_n|j_n}$,
where $p^\lambda_{k_i|j_i}$ is a vector of length $\sum_{j=1}^{m_i}v_{ij}$ such that
\begin{equation}\label{defdetvect3}
p^\lambda_{k_i|j_i}=\left\{\begin{array}{ll}1 &\text{if } \lambda_{ij_i}=k_i\\ 0&\mbox{otherwise .}\end{array}\right.
\end{equation}

For fixed $i$, consider, for each $j_i'\in\{1,\ldots,m_i\}$ and for each $k_i'\in\{2,\ldots,v_{ij_i'}\}$, the points $p^\lambda_{k_i|j_i}$ defined by $\lambda_{ij_i}=1$ for all $j_i\neq j_i'$ and $\lambda_{ij'_i}=k'_i$. In addition, consider the vector $p^\lambda_{k_i|j_i}$ defined by $\lambda_{ij_i}=1$ for all $j_i$. These $\sum_{j=1}^{m_i}(v_{ij}-1)+1$ points are linearly independent. The products $p^\lambda_{k_1\ldots k_n|j_1\ldots j_n}=p^\lambda_{k_1|j_1}\ldots p^\lambda_{k_n|j_n}$  of all these points thus define $\prod_{i=1}^{n}\Big(\sum_{j=1}^{m_i}(v_{ij}-1)+1\Big)=D+1$ linearly independent extreme points of $\mathcal{B}$, which are therefore also affinely independent.\hfill$\square$

Since $\mathcal{B}$ is not full dimensional, it follows that there is no unique way to write down a valid inequality for $\mathcal{B}$. More specifically, the inequalities $b\cdot p\geq b_0$ and $(b+\mu c)\cdot p\geq (b+\mu c_0)$, where $\mu \in \mathbb{R}$ and where $c\cdot p=c_0$ is a linear combination of the equalities (\ref{multinorma}) and (\ref{multinosig}), impose the same constraints on $\mathcal{B}$. In particular, it is always possible to use the normalization conditions to rewrite an inequality such that its lower bound is $0$, that is, in the form $b\cdot p\geq 0$. This fact will be used later on. 

\subsection{Trivial facets and nontrivial polytopes}\label{trivfac}
In addition to the normalization and nosignaling conditions, $\mathcal{B}$ also satisfy the following \emph{positivity conditions}:
\begin{equation}\label{multipos}
p_{k_1 \ldots k_n|j_1 \ldots j_n}\geq 0
\end{equation}
for all $k_1,\ldots,k_n$ and $j_1,\ldots,j_n$.
\begin{theorem}
The positivity conditions support facets of $\mathcal{B}$.
\end{theorem}
\noindent\emph{Proof.}
Without loss of generality, suppose that $p_{k_1 \ldots k_n|j_1 \ldots j_n}\geq 0$ is such that the $k_1,\ldots,k_n$ are all different than $1$. Then, in the proof of Theorem 1, we enumerated $\dim \mathcal{B}+1$ affinely independent points, $\dim \mathcal{B}$ of which satisfy $p_{k_1 \ldots k_n|j_1 \ldots j_n}=0$.\hfill$\square$

The normalization, nosignaling, and positivity conditions are obviously not only satisfied by local probabilities, but also by all nosignaling nonlocal ones, and in particular by quantum ones. The only useful constraints that separate the local region from the nonlocal thus correspond to the facets of $\mathcal{B}$ that are not of the form (\ref{multipos}).

Let us also note that when determining the facets of a Bell polytope, we can always assume that $n$, $m_i$ and $v_{ij}$ are all $\geq 2$ because otherwise all the corresponding facets are trivial or belong to simpler polytopes. Indeed,
\begin{enumerate}
\item[(i)] the only facet inequalities of one-partite polytopes are the positivity constraints,
\item[(ii)] all the facet inequalities of a polytope where $m_i=1$ for some party $i$ are equivalent to the facet inequalities of the polytope obtained by discarding that party,
\item[(iii)] a polytope with $v_{ij}=1$ for some measurement $j$ of party $i$ is equivalent to the polytope obtained by discarding that measurement choice.
\end{enumerate}
Point (i) is easily established. To show (ii), assume that $\mathcal{B}$ is a polytope such that for party $i$ the only measurement choice is $j\in\{1\}$. A valid inequality for $\mathcal{B}$ can thus be written as
\begin{equation}\label{trivmeas1}
\sum_k b_k\cdot p(i,j,k) \geq 0\,,
\end{equation}
where, without loss of generality, the right-hand side is equal to zero. It then follows that for all $k\in\{1,\ldots,v_{ij}\}$ the following inequalities
\begin{equation}\label{trivmeas2}
b_k\cdot p(i,j,k)\geq 0
\end{equation}
are also valid for $\mathcal{B}$. Indeed, for each extreme point $p^\lambda$, either the assignment $\lambda$ is such that $\lambda_{ij}=k$ and (\ref{trivmeas1}) and (\ref{trivmeas2}) impose the same constraints on $p^\lambda$, or $\lambda_{ij}\neq k$ and (\ref{trivmeas2}) gives the trivial inequality $0\geq 0$. Every extreme point satisfying (\ref{trivmeas1}) thus also satisfies (\ref{trivmeas2}). Note further that every extreme point satisfying (\ref{trivmeas1}) with equality also satisfies (\ref{trivmeas2}) with equality. This implies that the face supported by (\ref{trivmeas1}) cannot be --- unless (\ref{trivmeas1}) is itself equivalent to one of the inequalities (\ref{trivmeas2}) --- a facet of $\mathcal{B}$, because it lies in the intersection of the faces supported by (\ref{trivmeas2}) and is therefore of dimension $<\dim\mathcal{B}-1$. We can thus assume that all facet inequalities of $\mathcal B$ are of the form (\ref{trivmeas2}). It will be shown in Section \ref{morobs}, that all these facet inequalities are equivalent to facet inequalities of the polytope obtained by discarding party $i$. Finally, point (iii) follows immediately when we notice that a polytope with $v_{ij}=1$ for some measurement $j$ of party $i$ and the polytope obtained by discarding that measurement have the same dimension and have their extreme points  in one-to-one correspondence. 

\subsection{A useful lemma}
As we have reminded earlier an inequality defines a facet of a polytope $\mathcal{B}$ if and only if it is satisfied by $\dim{B}$ affinely independent points of $\mathcal{B}$. To prove the results of the next section concerning the lifting of facet inequalities, we will then need to count the number of affinely independent points that a facet contains. The following lemma will be our main tool to achieve this task.

\begin{lemma}\label{lemma}
Let the inequality $b\cdot p\geq b_0$ support a facet of $\mathcal{B}(n,m,v)$. Let $i'\in\{1,\ldots,n\}$, $j'\in\{1,\ldots,m_{i'}\}$ and $k'\in\{1,\ldots,v_{i'j'}\}$. Then there are at exactly $r$ extreme points $p^\lambda$ of $\mathcal{B}$ such that $b\cdot p^\lambda=b_0$, $\lambda_{i'j'}=k'$, and such that the $r$ restrictions $p^\lambda(i',j',k')$ are affinely independent, where 
\begin{enumerate}
\item[(i)] $r=\prod_{i\neq i'}\big(\sum_{j=1}^{m_i}(v_{ij}-1)+1\big)-1$, if $b\cdot p\geq b_0$ is equivalent to an inequality of the form\linebreak[4] $c\cdot p(i',j',k')\geq 0$;
\item[(ii)]  $r=\prod_{i\neq i'}\big(\sum_{j=1}^{m_i}(v_{ij}-1)+1\big)$, otherwise.
\end{enumerate}
\end{lemma}
\noindent\emph{Proof.}
Let $\{p^\delta\,|\,\delta\in \Delta\subseteq\Lambda\}$ be $\dim\mathcal{B}$ affinely independent extreme points which belong to the facet supported by $b\cdot p\geq b_0$. Among these, let $\{p^\gamma\,|\,\gamma\in \Gamma\subseteq\Delta\}$ be the extreme points satisfying $\gamma_{i'j'}=k'$ and such that their restrictions $\{p^\gamma(i',j',k')\,|\gamma\in \Gamma\}$ are affinely independent.

Consider the polytope $\mathcal{B}^{n-1}$ obtained from $\mathcal{B}$ by discarding party $i'$. The components of  $p\in\mathcal{B}^{n-1}$ are thus of the form $p_{k_1\ldots k_{i'-1}k_{i'+1}\ldots k_n|j_1\ldots j_{i'-1}j_{i'+1}\ldots j_n}$. Given that $p^\gamma(i',j',k')$ corresponds to the components of $p^\gamma$ where the indices associated to the ${i'}^\mathrm{th}$ party are fixed and satisfy $k_{i'}=k'$, $j_{i'}=j'$, given that $\gamma_{i'j'}=k'$, and given definition (\ref{defdetvect}), it follows that each $p^\gamma(i',j',k')$ can be identified with an extreme point of the $(n-1)$-partite polytope $\mathcal{B}^{n-1}$ (and conversely, each extreme point of $\mathcal{B}^{n-1}$ can be identified with the restriction $p^\gamma(i',j',k')$ of some extreme point $p^\gamma\in\mathcal{B}$ satisfying $\gamma_{i'j'}=k'$). Thus no more than $\dim\mathcal{B}^{n-1}$ of the $p^\gamma(i',j',k')$ can be affinely independent, and $r\leq\dim\mathcal{B}^{n-1}+1=\prod_{i\neq i'}\big(\sum_{i=1}^{m_i}(v_{ij}-1)+1\big)$. Alternatively, one could have deduced the same result starting from the fact that the $p^\gamma$ satisfy the implicit equalities (\ref{multinorma}) and (\ref{multinosig}), and counting the number of constraints that these equalities impose on the $p^\gamma(i',j',k')$.

Suppose that $r<\dim\mathcal{B}^{n-1}+1$. Then the $\{p^\gamma\,|\,\gamma\in \Gamma\}$ satisfy at least one constraint
\begin{equation}\label{dmn}
c\cdot p(i',j',k')=0
\end{equation}
linearly independent from the implicit equalities of $\mathcal{B}$. Following the remark at the end of Section \ref{dim}, we have not lost generality by taking the right-hand side of (\ref{dmn}) equal to zero. Note that the constraint (\ref{dmn}) is in fact satisfied by all $\{p^\delta\,|\,\delta\in \Delta\}$. Indeed, either $\delta_{i'j'}\neq k'$ and (\ref{dmn}) gives the trivial equation $0=0$, or $p^\delta(i',j',k')$ is affinely dependent from the $p^\gamma(i',j',k')$, which satisfy (\ref{dmn}).

As the $\{p^\delta\,|\,\delta\in \Delta\}$ form a set of $\dim\mathcal{B}$ independent extreme points, they can satisfy at most one constraint linearly independent from the implicit equalities of $\mathcal{B}$, i.e., there can only be one constraint of the form (\ref{dmn}). Thus at most $r=\dim\mathcal{B}^{n-1}=\prod_{i\neq i'}\big(\sum_{i=1}^{m_i}(v_{ij}-1)+1\big)-1$. Furthermore, as the $\{p^\delta\,|\,\delta\in \Delta\}$ already satisfy the equality $b\cdot p=b_0$, this can only be the case if (\ref{dmn}) is equivalent to $b\cdot p=b_0$, that is if $b\cdot p\geq b_0$ is equivalent either to $c\cdot p(i',j',k')\geq 0$ or $(-c)\cdot p(i',j',k')\geq 0$. \hfill$\square$

\section{Lifting Bell inequalities}\label{sectlifting}
We now move on to study the liftings of Bell inequalities that we have presented in the introduction and their natural generalizations. We will prove that these liftings are facet-preserving. It was already shown in \cite{aii04} that a Bell inequality that supports a facet of $\mathcal{B}(2,m,2)$ also supports a facet of $\mathcal{B}(2,m',2)$ for all $m'\geq m$. Furthermore, in \cite{kvk98} liftings of ``partial constraint satisfaction polytopes" (polytopes encountered in certain optimization problems) were considered. Although such liftings were studied independently from any potential relation to Bell inequalities, it turns out that partial constraint satisfaction polytopes over a complete bipartite graph are bipartite Bell polytopes (in particular, the ``4-cycle inequality" introduced in \cite{kvk98} corresponds to the CHSH inequality). The results presented in \cite{kvk98} then imply that an inequality that supports a facet of $\mathcal{B}(2,m,v)$ also supports a facet of $\mathcal{B}(2,m',v')$ for all $m'\geq m$, $v'\geq v$. It is in fact these results that inspired the ones that are presented here. 

In the next three subsections, we will see that the lifting of an arbitrary inequality to a situation involving, respectively, one more observer, one more measurement outcome, and one more measurement setting are facet-preserving. Combined together these results imply that a Bell inequality that supports a facet of a Bell polytope $\mathcal{B}(n,m,v)$, also supports, when lifted in the appropriate way, a facet of any higher dimensional polytope $\mathcal{B}(n',m',v')$ with $n'\geq n$, $m'\geq m$, $v'\geq v$.

\subsection{One more observer}\label{morobs}
Consider a polytope $\mathcal{B}\equiv\mathcal{B}(n,m,v)$, where the $n$ parties are labeled $\{1,\ldots,i'-1,i'+1\ldots,n+1\}$ for some value $i'$. Let the inequality 
\begin{equation}\label{origineqpart}
b\cdot p\geq 0
\end{equation}
be valid for $\mathcal{B}$. Note that we have taken, without loss of generality, the right-hand side of (\ref{origineqpart}) to be equal to $0$. Let us extend the polytope $\mathcal{B}$ by inserting an additional observer in position $i'$. The resulting $(n+1)$-partite polytope will be denoted $\mathcal{B}^{n+1}$. 

Given a point $p\in\mathcal{B}^{n+1}$, remember that $p(i',j',k')$ represents the probabilities of $p$ for which the indices corresponding to the measurement setting and the outcome of party $i'$ are fixed, and are equal, respectively, to $j'$ and $k'$. Therefore $p(i',j',k')/p_{k'_{i'}|j'_{i'}}$, where $p_{k'_{i'}|j'_{i'}}$ denotes the marginal probability for observer $i'$ to measure $j'$ and obtain $k'$, is the joint outcome probability distribution for the $n$ observers $\{1,\ldots,i'-1,i'+1,\ldots n+1\}$ conditional on party $i'$ measuring $j'$ and obtaining $k'$. Either this conditional probability is equal to zero, or it corresponds to a point of $\mathcal{B}$. In both cases, it satisfies (\ref{origineqpart}). It thus follows immediately that the following inequality
\begin{equation}\label{liftineqpart}
b\cdot p(i',j',k')\geq 0
\end{equation}
is valid for $\mathcal{B}^{n+1}$. Further, this lifting is facet-preserving.

\begin{theorem}\label{liftparttheo}
The inequality (\ref{origineqpart}) supports a facet of $\mathcal{B}$ if and only if (\ref{liftineqpart}) supports a facet of $\mathcal{B}^{n+1}$.
\end{theorem}
\noindent\emph{Proof.}
As we have noted in the proof of Lemma \ref{lemma}, the restriction $p^\lambda(i',j',k')$ of an extreme point $p^\lambda$ of $\mathcal{B}^{n+1}$ satisfying $\lambda_{i'j'}=k'$ can be identified with an extreme point of $\mathcal{B}$, and conversely. Moreover, it is clear that if $p^\lambda(i',j',k')$ satisfy (\ref{liftineqpart}) with equality the corresponding extreme point of $\mathcal{B}$ satisfy (\ref{origineqpart}) with equality, and the other way around.

Assume that (\ref{liftineqpart}) supports a facet of $\mathcal{B}^{n+1}$. Then it follows from Lemma \ref{lemma} that they are $\prod_{i\neq i'}\big(\sum_{j=1}^{m_i}(v_{ij}-1)+1\big)-1=\dim \mathcal{B}$ extreme points of $\mathcal{B}^{n+1}$ that satisfy (\ref{liftineqpart}) with equality, such that $\lambda_{i'j'}=k'$ and for which the restrictions $p^\lambda(i',j',k')$ are affinely independent. By the above remark, these extreme points define $\dim \mathcal{B}$ affinely independent extreme points of $\mathcal{B}$ that satisfy (\ref{origineqpart}) with equality, hence this inequality supports a facet of $\mathcal{B}$. 

To prove the converse statement, suppose now that (\ref{origineqpart}) defines a facet of $\mathcal{B}$, that is, there exist $\dim \mathcal{B}$ affinely independent extreme points of $\mathcal{B}$ that satisfy it with equality. By the above remark, there thus exist $\dim\mathcal{B}$ extreme points of $\mathcal{B}^{n+1}$ that satisfy (\ref{liftineqpart}) with equality, such that $\lambda_{i'j'}=k'$ and for which the restrictions $p^\lambda(i',j',k')$ are affinely independent. To show that (\ref{liftineqpart}) defines a facet of $\mathcal{B}^{n+1}$, it thus remain to find $\dim\mathcal{B}^{n+1}-\dim\mathcal{B}$ affinely independent points satisfying it with equality. For this, consider\footnote{We use the fact that $v_{ij}\geq 2$, following the remark at the end of Section \ref{trivfac}.} the extreme points of $\mathcal{B}^{n+1}$ with $\lambda_{i'j'}\neq k'$. They form an affine subspace of dimension $\dim\mathcal{B}^{n+1}-\prod_{i\neq i'}\big(\sum_{j=1}^{m_i}(v_{ij}-1)+1\big)=\dim\mathcal{B}^{n+1}-\dim\mathcal{B}-1$ since they can be identified with the extreme points of the polytope involving one outcome less than $\mathcal{B}^{n+1}$ for the measurement $j'$. Moreover, because they verify $p^\lambda(i',j',k')=0$, they satisfy (\ref{liftineqpart}) with equality, and are affinely independent from the extreme points for which $\lambda_{i'j'}=k'$.
\hfill$\square$

We thus have just shown that any facet inequality of an $n$-partite polytope can be extended to a facet inequality for a situation involving $n+1$ parties. This result can be used sequentially so that facets of $n$-party polytopes are lifted to $(n+k)$-partite polytopes. For instance, the positivity conditions (\ref{multipos}) can be viewed as the successive lifting of $1$-party inequalities. 

The result holds in the other direction as well, since any facet inequality of the form (\ref{liftineqpart}) is the lifting of an $n$-partite inequality. When studying Bell polytopes, it is thus in general sufficient to consider \emph{genuinely $n$-partite inequalities}, that is, inequalities that cannot be written in a form that involves only probabilities associated with one specific measurement setting $j'$ and one specific outcome $k'$ for some party $i'$.  Note that we can extend this definition to exclude also all inequalities such as (\ref{trivmeas1}) that involve only probabilities associated to one measurement setting (but possibly several outcomes corresponding to this measurement). Indeed, we have noted at the end of section \ref{trivfac} that such inequalities cannot be stronger than inequalities of the form (\ref{liftineqpart}). 

\subsection{One more measurement outcome}
Consider a polytope $\mathcal{B}\equiv\mathcal{B}(n,m,v)$, where for measurement $j'$ of party $i'$ the $v_{i'j'}$ outcomes are labeled $\{1,\ldots,k'-1,k'+1,\ldots,v_{i'j'}+1\}$ for some $k'$. Let
\begin{equation}\label{origineqpart2}
b\cdot p\geq b_0
\end{equation}
be a genuinely $n$-partite inequality valid for $\mathcal{B}$. Let us consider the polytope $\mathcal{B}^{v+1}$ obtained from $\mathcal{B}$ by allowing an extra outcome $k'$ for the measurement $j'$ of party $i'$. To lift the inequality $b\cdot p\geq b_0$ to the polytope $\mathcal{B}^{v+1}$, we can merge the additional outcome $k'$ with some other outcome $k^*\in\{1,\ldots,k'-1,k'+1,\ldots,v_{i'j'}+1\}$, and insert the resulting probability distribution in (\ref{origineqpart}). This results in the inequality
\begin{equation}\label{liftineqout}
b\cdot p+b(i',j',k^*)\cdot p(i',j',k')\geq b_0\,.
\end{equation}
\begin{theorem}\label{theoliftout}
If the genuinely $n$-partite inequality (\ref{origineqpart}) supports a facet of $\mathcal{B}$, then (\ref{liftineqout}) supports a facet of $\mathcal{B}^{v+1}$.
\end{theorem}
\noindent\emph{Proof.}
The dimension of $\mathcal{B}^{v+1}$ equals $\dim\mathcal{B}+\prod_{i\neq i'}\big(\sum_{j=1}^{m_i}(v_{ij}-1)+1\big)$. The extreme points of $\mathcal{B}$ that belong to the facet $b\cdot p\geq b_0$ provide $\dim\mathcal{B}$ affinely independent points satisfying (\ref{liftineqout}) with equality. By Lemma \ref{lemma}, there exist $\prod_{i\neq i'}\big(\sum_{j=1}^{m_i}(v_{ij}-1)+1\big)$ extreme points $p^\lambda$ with $\lambda_{i'j'}=k^*$ that saturate (\ref{origineqpart}), and thus (\ref{liftineqout}), and for which the $p^\lambda(i',j',k^*)$ are affinely independent. Replace $k^*$ by $k'$ in these extreme points. These new extreme points still satisfy (\ref{liftineqout}) with equality and are affinely independent with all the previous ones, since they are the unique extreme points with $p^\lambda(i',j',k')\neq 0$. In total, we thus enumerated $\dim\mathcal{B}^{v+1}=\dim\mathcal{B}+\prod_{i\neq i'}\big(\sum_{j=1}^{m_i}(v_{ij}-1)+1\big)$ affinely independent point satisfying (\ref{liftineqout}) with equality.
\hfill$\square$

\subsection{One more measurement setting}
Consider a polytope $\mathcal{B}\equiv\mathcal{B}(n,m,v)$, where for party $i'$ the $m_{i'}$ measurements are labeled $\{1,\ldots,j'-1,j'+1,\ldots,m_{i'}+1\}$ for some $j'$. Let the polytope $\mathcal{B}^{m+1}$ be the polytope obtained from $\mathcal{B}$ by allowing the additional measurement setting $j'$ for party $i'$. An inequality $b\cdot p\geq b_0$ valid for $\mathcal{B}$ is also clearly valid for $\mathcal{B}^{m+1}$. Moreover, the following stronger result holds.
\begin{theorem}
Let $b\cdot p\geq b_0$ be a genuinely $n$-partite inequality supporting a facet of $\mathcal{B}$. Then it is also support a facet of $\mathcal{B}^{m+1}$.
\end{theorem}
\noindent\emph{Proof.}
Consider the polytope $\widetilde{\mathcal{B}}^{m+1}$ defined as $\mathcal{B}^{m+1}$ but such that for the measurement $j'$ of party $i'$ is associated a single possible outcome, i.e., $v_{i'j'}=1$. The inequality $b\cdot p\geq b_0$ is a valid genuinely $n$-partite inequality for $\widetilde{\mathcal{B}}^{m+1}$. Further, since $\widetilde{\mathcal{B}}^{m+1}$ and $\mathcal{B}$ have the same dimension, it is also facet defining for $\widetilde{\mathcal{B}}^{m+1}$. Following the procedure to lift an inequality to more outcomes delineated in the previous subsection, this inequality can be lifted from $\widetilde{\mathcal{B}}^{m+1}$ to $\mathcal{B}^{m+1}$. Since $b\cdot p\geq b_0$ does not involve components associated with the measurement $j'$ of party $i'$, this results in the inequality $b\cdot p\geq b_0$ itself. By Theorem \ref{theoliftout}, this inequality is facet defining for $\mathcal{B}^{m+1}$. 
\hfill$\square$

\section{Conclusion}
We have shown that the facial structure of Bell polytopes is organized in a hierarchical way, with all the facets of a given polytope inducing, through their respective liftings, facets of more complex polytopes. Instead of considering the entire set of facets of a Bell polytope, it is thus in general sufficient to characterize the ones that do not belong to simpler polytopes. It would be interesting to investigate whether this fact could be exploited to improve the efficiency of the algorithms used to list facet inequalities or to simplify analytical derivations of Bell inequalities.

Note that for certain polytopes, the complete set of facet inequalities is constituted entirely by inequalities lifted from more elementary polytopes. For instance for Bell scenarios involving two observers, the first having a choice between two dichotomic measurements and the second one between an arbitrary number of them, all the facet-defining inequalities correspond to liftings of the CHSH inequality \cite{sli03,cg04}. A natural extension of the results reported in this article would then be to investigate more generally when inequalities lifted from simpler polytopes describe complete sets of facets. Progress along this line would allow one to narrow down the class of Bell scenarios that have to be considered to find new Bell inequalities.  
Following this approach, all the polytopes for which the only facets correspond to liftings of the CHSH inequality have recently been characterized \cite{sp}. 

Finally, let us note that while the facet-preserving liftings that we have considered are interesting because they throw light on the structure of Bell polytopes, the inequalities obtained in this way are not essentially different from the original ones, they are merely re-expressions of these inequalities adapted to more general scenarios. However, it is also in principle possible to consider more complicated generalizations of Bell inequalities that alter significantly their intrinsic structure. For instance, the family of Bell inequalities introduced in \cite{cgl02} can be understood as being generated by successive nontrivial liftings of the CHSH inequality. Studying such liftings, as well as the other possible extensions of our results, seems a promising path towards a more accurate characterization of the constraints that separate the set of local joint probabilities from the set of nonlocal ones.

\acknowledgments
I would like to thank Jean-Paul Doignon and Serge Massar for helpful discussions. This work is supported by the David and Alice Van Buuren fellowship of the Belgian American Educational Foundation and by the National Science Foundation under Grant No. EIA-0086038.

\end{document}